\documentstyle[preprint,eqsecnum,aps]{revtex}

\begin{document}
\draft
\preprint{Preprint-INPP-UVA-93-1}

\title{Sea Contributions and Nucleon Structure}
\author{X. Song}

\address{Institute of Nuclear and Particle Physics,
Department of Physics,\\
University of Virginia, Charlottesville, VA 22901, USA}

\author{V. Gupta}

\address{Tata Institute of Fundamental Research, Bombay 400005, India}

\maketitle
\begin{abstract}
We suggest a general formalism to treat a baryon as a composite
system of three quarks and a `sea'. In this formalism, the sea
is a cluster which can consists of gluons and quark-antiquark
pairs. The hadron wave function with a sea component is given.
The magnetic moments, related sum rules and axial weak coupling
constants are obtained. The data seems to favor a vector sea
rather than a scalar sea. The quark spin distributions in the
nucleon are also discussed.
\end{abstract}

\widetext

\section{INTRODUCTION}

Historically, the static SU(6) quark model provided a
good description of hadrons: Baryons (mesons) are
color-singlet combinations of three quarks (quark
antiquark pairs) in the appropriate flavor and spin
combination. The space-time part of a hadron wave
function can be determined by using a specific model
of confinement, {\it e.g.} bag model\cite{chodos74a,chodos74b}
simple harmonic oscillator model\cite{isgur79,godfery85,capstick86},
or other phenomenological models\cite{lichtenberg87}.
Although the naive SU(6) quark model works successfully
in explaining various properties of hadrons, departures
from the naive SU(6) results have been observed. The
naive $valence$ picture of hadron structure is a
simplification or a first order approximation to the
real system. Within the framework of QCD, quarks
interact through color forces mediated by vector gluons.
The QCD interaction Hamiltonian
$H_I(x)=g{\bar{\psi}}(x){\gamma}^{\mu}({\lambda}^a/2){\psi}
(x)A_{\mu}^a(x) $
has several consequences: First of all, spin-dependent
forces (e.g. color-hyperfine interactions\cite{rujula75})
between the quarks due to one (or multi-) gluon exchange
lift the SU(6) mass degeneracy and explain the basic
pattern of baryon and meson spectroscopies. The spin
dependent forces also cause different space-time
distributions for different quark flavors and provide
a good description of baryon magnetic moments and form
factors\cite{isgur81,song92}. Secondly, the existence
of quark-gluon interaction implies that quark-antiquark
($q\bar q$-)pairs can be created by the virtual gluons
emitted from valence quarks. These $q\bar q$-pairs are
the so called sea quarks. Usually, the `$sea$' means a
combination of the virtual gluons and sea quark-antiquark
pairs. Although deep inelastic muon nucleon scattering
shows that the sea components ($q\bar q$-pairs and gluons)
indeed exist and play a very important role ({\it e.g.}
gluons carry about one half of the nucleon momentum and
the $sea$ dominates small$-x$ behaviour of structure
functions), it is commonly believed that in the low
energy regime, static properties of hadrons are dominated
by their valence components. However, it has been shown
\cite{donoghue77,he84} that the sea contributions may
change the structure of hadrons and modify their low
energy properties. Using the QCD interaction Hamiltonian
and the MIT bag model, Donoghue and Golowich (DG)\cite
{donoghue77} (comments see cf \cite{he84}) calculated the
probabilities of different sea quark components in the
proton. Several models\cite{golowich83,close90,close88,li91}
have been suggested to study the gluon component in hadrons.
In these models, a mixing of $q^3$ and $q^3$+gluon, in which
a color ${\bf 8}_c$ gluon coupled to a ${\bf 8}_c$ $q^3$
state to form a color singlet, has been discussed. However,
the ``sea'' could be a gluon (as discussed in
\cite{golowich83,close90,close88,li91}) or a quark-antiquark
pair (as discussed in \cite{donoghue77,he84}), or
even more complicated, for instance a multi-gluon state,
multi-($q\bar q$-) pairs or gluon(s) plus ($q\bar q$-)
pair(s). In this paper, we study the sea contributions
in a more general formalism and treat the ``sea'' as a
cluster which can consist of two-gluon and a gluon
plus a ($q-\bar q$) pair or some admixture of both
(which may be described by the generic term ``flotsam'').
Since the baryon should be colorless and a $q^3$ state
can be in color states ${\bf 1}_c$, ${\bf 8}_c$, and
${\bf 10}_c$, the ``sea'' should also be in corresponding
color states to form a color singlet baryon. In addition,
the ``sea'' spin is not required to be one (as in the
single-gluon case). Furthermore, if the sea is in a
S-wave state relative to the $q^3$ system, conservation
of the angular momentum restricts that sea spin can only
be 0, 1 or 2 to give a spin-1/2 baryon. If the sea is in
a P-wave state, then its spin could be 0, 1, 2, or 3. In
this paper, we only discuss the S-wave case. In section
II, a more general wave function of the baryon, which
consists of $q^3$ and a ``sea'', is given. In section
III, the magnetic moments and related sum rules are
derived and compared with the data. In section IV, axial
weak coupling constants and first moments of nucleon spin
structure functions are calculated. A discussion of the
sea contribution, numerical results and several conclusions
are given in section V, VI and VII respectively.

\section{HADRON WAVE FUNCTION WITH A SEA COMPONENT.}

The three (valence) quark wave function of the baryon
can be written as

\begin{equation}
$$ \Psi =\Phi (| \phi>\cdot | \chi>\cdot |
\psi>)\cdot (| \xi>)          $$
\end{equation}
where $| \phi>$, $| \chi>$, $| \psi>$ and
$| \xi>$ denote flavor, spin, color and space-time
$q^3$ wave functions. For the lowest-lying hadrons,
quarks appear to be in S-wave states and the
space-time $q^3$ wave function $| \xi>$ is total
symmetric under permutation of any two quarks. Hence
the flavor-spin-color part $\Phi$ should be total
antisymmetric under $q_i\leftrightarrow q_j$. In the
conventional quark model, the color wave function
$\psi$ is taken to be total antisymmetric, i.e. a
color singlet. But in general this is not necessary
if baryon is considered to have a sea component in
addition to the $q^3$. Let superscripts $S$ and $A$
denote total permutation symmetry and antisymmetry, and
$\lambda$, $\rho$ denote symmetric and antisymmetric
under quark permutation $q_1\leftrightarrow q_2$. Then
the $q^3$ wave functions for a flavour octet baryon, are

\begin{equation}
$${\Phi}_1^{(1/2)}\equiv \Phi ({\bf 8},1/2,{\bf 1}_c)
=F_S{\psi}_{1}^A                       $$
\end{equation}

\begin{equation}
$${\Phi}_8^{(1/2)}\equiv \Phi ({\bf 8}, 1/2, {\bf 8}_c)
={1\over {\sqrt 2}}(F_{MS}
{\psi}_8^{\rho}-F_{MA}{\psi}_{8}^{\lambda})  $$
\end{equation}

\begin{equation}
$${\Phi}_{10}^{(1/2)}\equiv \Phi ({\bf 8},1/2,{\bf 10}_c)
=F_A{\psi}_{10}^S                            $$
\end{equation}

\begin{equation}
$${\Phi}_8^{(3/2)}\equiv \Phi ({\bf 8},3/2,{\bf 8}_c)
=F_A'{\chi}^{(3/2)}                        $$
\end{equation}
where

\begin{equation}
$$F_S={1\over {\sqrt 2}}({\phi}^{\lambda}{\chi}^{\lambda}+
{\phi}^{\rho}{\chi}^{\rho})                  $$
\end{equation}

\begin{equation}
$$F_{MS}={1\over {\sqrt 2}}({\phi}^{\rho}{\chi}^{\rho}-
{\phi}^{\lambda}{\chi}^{\lambda})           $$
\end{equation}

\begin{equation}
$$F_{MA}={1\over {\sqrt 2}}({\phi}^{\rho}{\chi}^{\lambda}+
{\phi}^{\lambda}{\chi}^{\rho})               $$
\end{equation}

\begin{equation}
$$F_{A}={1\over {\sqrt 2}}({\phi}^{\lambda}{\chi}^{\rho}-
{\phi}^{\rho}{\chi}^{\lambda})                $$
\end{equation}
and

\begin{equation}
$$F_{A}'={1\over {\sqrt 2}}({\phi}^{\lambda}{\psi}_8^{\rho}-
{\phi}^{\rho}{\psi}_8^{\lambda})              $$
\end{equation}
where the detail expressions for ${\phi}^{\lambda}$,
${\phi}^{\rho}$, ${\chi}^{\lambda}$ and ${\chi}^{\rho}$
can be found in Ref.\cite{close79}, and ${\chi}^{(3/2)}$
is the totally symmetric $q^3$ spin wave function with
spin 3/2.

We note that ${\Phi}_1^{(1/2)}$ in (2.2) is the standard $q^3$
wave function which transforms as $\bf 56$ of SU(6) and was
denoted by $| N_0>$ in Ref.\cite{li91}. Our ${\Phi}_8^{(1/2)}$ and
${\Phi}_8^{(3/2)}$ correspond to the notation $\mid ^2N_g>$
and $|^4N_g>$ in Ref.\cite{li91} respectively, they transform as
$\bf 70$ of SU(6). There is no ${\Phi}_{10}^{(1/2)}$ term
in previous works.

We consider a flavorless sea, which has spin (0,1,2 if we
assume sea is in a S wave state) and color (${\bf 1}_c$,
${\bf 8}_c$ and ${\bf {\bar 10}}_c$,). Let
$H_{0,1,2}$ and $G_{1,8,{\bar 10}}$ denote spin
and color sea wave functions, which satisfy

\begin{equation}
$$<H_i| H_j>={\delta}_{ij}\quad,\qquad
<G_k| G_l>={\delta}_{kl}       $$
\end{equation}
The possible combinations of $q^3$ and sea wave functions,
which can give a spin 1/2, flavour octet, color singlet state,
are:
\begin{equation}
$${\Phi}_1^{(1/2)}\cdot H_0\cdot G_1\ ,\quad
 {\Phi}_8^{(1/2)}\cdot H_0\cdot G_8\ ,\quad
{\Phi}_{10}^{(1/2)}\cdot H_0\cdot G_{{\bar 10}}    $$
\end{equation}

\begin{equation}
$${\Phi}_1^{(1/2)}\cdot H_1\cdot G_1\ ,\quad
 {\Phi}_8^{(1/2)}\cdot H_1\cdot G_8\ ,\quad
{\Phi}_{10}^{(1/2)}\cdot H_1\cdot G_{{\bar 10}}    $$
\end{equation}
and

\begin{equation}
$${\Phi}_8^{(3/2)}\cdot H_1\cdot G_8\ ,\quad
{\Phi}_8^{(3/2)}\cdot H_2\cdot G_8               $$
\end{equation}
The total flavor-spin-color wave function of a spin up
baryon which consists of three valence quarks and a sea
component can be written as

\begin{eqnarray}
| {\Phi}_{1/2}^{({\uparrow})}>
&=&{1\over N}\Bigl [ {\Phi}_1^{(1/2{\uparrow})}\cdot H_0\cdot G_1+
a_8{\Phi}_8^{(1/2{\uparrow})}\cdot H_0\cdot G_8+
a_{10}{\Phi}_{10}^{(1/2{\uparrow})}\cdot H_0\cdot G_{{\bar 10}}\nonumber \\
&&+b_1({\Phi}_1^{(1/2)}\otimes H_1)^{\uparrow}\cdot G_1+
b_8({\Phi}_8^{(1/2)}\otimes H_1)^{\uparrow}\cdot G_8+
b_{10}({\Phi}_{10}^{(1/2)}\otimes H_1)^{\uparrow}\cdot G_{{\bar 10}}\nonumber
\\
&&+c_8({\Phi}_8^{(3/2)}\otimes H_1)^{\uparrow}\cdot G_8+
d_8({\Phi}_8^{(3/2)}\otimes H_2)^{\uparrow}\cdot G_8\Bigr ]
\end{eqnarray}
where

\begin{equation}
$$N^2=1+a_8^2+a_{10}^2+b_1^2+b_8^2+b_{10}^2+c_8^2+d_8^2   $$
\end{equation}
Although there are seven correction terms in (2.15), they
are not equally important. Some arguments are given in
section V to show that main modifications come from the
vector sea, in particular $b_8$, $b_1$ and $c_8$ terms,
and minor contributions come from the scalar sea, e.g.
$a_{10}$ term.

The first three terms in (2.15) come from a spin 1/2 $q^3$
state coupled to a spin 0 (scalar) sea. The next three
terms in (2.15) come from spin 1/2 $q^3$ $\otimes$
spin 1 (vector) sea and in more detail we have

\begin{equation}
$$({\Phi}_1^{(1/2)}\otimes H_1)^{\uparrow}\equiv
{\Phi}_{b1}^{(1/2{\uparrow})}{\psi}_1^A      $$
\end{equation}

\begin{equation}
$$({\Phi}_8^{(1/2)}\otimes H_1)^{\uparrow}\equiv
{\Phi}_{b8}^{(1/2{\uparrow})} $$
\end{equation}

\begin{equation}
$$({\Phi}_{10}^{(1/2)}\otimes H_1)^{\uparrow}\equiv
{\Phi}_{b10}^{(1/2{\uparrow})}{\psi}_{10}^S      $$
\end{equation}
where

\begin{equation}
$${\Phi}_{b1}^{(1/2{\uparrow})}={\sqrt {2\over 3}}
H_{1,1}F_S^{(1/2{\downarrow})}
-{\sqrt {1\over 3}}H_{1,0}F_S^{(1/2{\uparrow})}  $$
\end{equation}

\begin{equation}
$${\Phi}_{b8}^{(1/2{\uparrow})}={\sqrt {1\over 2}}
[{\Phi}_{b8S}^{(1/2{\uparrow})}{\psi}_8^{\rho}
-{\Phi}_{b8A}^{(1/2{\uparrow})}{\psi}_8^{\lambda}]$$
\end{equation}

\begin{equation}
$${\Phi}_{b10}^{(1/2{\uparrow})}={\sqrt {2\over 3}}
H_{1,1}F_{A}^{(1/2{\downarrow})}
-{\sqrt {1\over 3}}H_{1,0}F_{A}^{(1/2{\uparrow})}$$
\end{equation}
In (2.21), ${\Phi}_{b8S}^{(1/2{\uparrow})}$ and
${\Phi}_{b8A}^{(1/2{\uparrow})}$ are

\begin{equation}
$${\Phi}_{b8S}^{(1/2{\uparrow})}={\sqrt {2\over 3}}
H_{1,1}F_{MS}^{(1/2{\downarrow})}
-{\sqrt {1\over 3}}H_{1,0}F_{MS}^{(1/2{\uparrow})}$$
\end{equation}

\begin{equation}
$${\Phi}_{b8A}^{(1/2{\uparrow})}={\sqrt {2\over 3}}
H_{1,1}F_{MA}^{(1/2{\downarrow})}
-{\sqrt {1\over 3}}H_{1,0}F_{MA}^{(1/2{\uparrow})}$$
\end{equation}
The final two ($c_8$, $d_8$) terms in Eq.(2.15) come from
spin 3/2 ($q^3$) $\otimes$ spin 1 (sea) and spin 3/2 ($q^3$)
$\otimes$ spin 2 (tensor sea) respectively. Their expressions
are

\begin{equation}
$$({\Phi}_8^{(3/2)}\otimes H_1)^{\uparrow}\equiv
{\Phi}_{c8}^{(1/2{\uparrow})}        $$
\end{equation}

\begin{equation}
$$({\Phi}_8^{(3/2)}\otimes H_2)^{\uparrow}\equiv
{\Phi}_{d8}^{(1/2{\uparrow})}         $$
\end{equation}
where

\begin{equation}
$${\Phi}_{c8}^{(1/2{\uparrow})}=[{1\over {\sqrt 2}}H_{1,-1}
{\chi}^{(3/2)}_{3/2}-{1\over {\sqrt 3}}H_{1,0}{\chi}^{(3/2)}_{1/2}
+{1\over {\sqrt 6}}H_{1,1}{\chi}^{(3/2)}_{-1/2}]F_A'  $$
\end{equation}

\begin{equation}
$${\Phi}_{d8}^{(1/2{\uparrow})}=[{\sqrt {2\over 5}}H_{2,2}
{\chi}^{(3/2)}_{-3/2}-{\sqrt {3\over 10}}H_{2,1}{\chi}^{(3/2)}_{-1/2}
+{\sqrt {1\over 5}}H_{2,0}{\chi}^{(3/2)}_{1/2}-
{\sqrt {1\over {10}}}H_{2,-1}{\chi}^{(3/2)}_{3/2}]F_A'$$
\end{equation}
The wave function used in Ref.\cite{li91} (see Eq.(3.9) in \cite{li91})
can be obtained from (2.15) by taking $a_{8,10}=b_{1,10}=d_8$=0
and $b_{8}$=$c_8$=$-\delta$. However, we would not like to
restrict ourself to this special case.

\section{Magnetic Moments and Related Sum Rules}

For any operator $\hat O$ which only depends on quark
flavor and spin and does not depend on the color and
space-time, we have

\begin{eqnarray}
<{\Phi}_{1/2}^{({\uparrow})}| {\hat O}|
{\Phi}_{1/2}^{({\uparrow})}>&=&
{1\over {N^2}}\Bigl [ <{\Phi}_1^{(1/2{\uparrow})}
| {\hat O}| {\Phi}_1^{(1/2{\uparrow})}>\nonumber \\
&&+\sum_{i=8,10}a_i^2
<{\Phi}_i^{(1/2{\uparrow})}
| {\hat O}| {\Phi}_i^{(1/2{\uparrow})}>\nonumber \\
&&+\sum_{i=1,8,10}b_i^2<{\Phi}_{bi}^{(1/2{\uparrow})}
| {\hat O}| {\Phi}_{bi}^{(1/2{\uparrow})}>\nonumber \\
&&+2b_8c_8<{\Phi}_{b8}^{(1/2{\uparrow})}| {\hat O}|
{\Phi}_{c8}^{(1/2{\uparrow})}>\nonumber \\
&&+c_8^2<{\Phi}_{c8}^{(1/2{\uparrow})}| {\hat O}|
{\Phi}_{c8}^{(1/2{\uparrow})}>\nonumber \\
&&+d_8^2<{\Phi}_{d8}^{(1/2{\uparrow})}| {\hat O}|
{\Phi}_{d8}^{(1/2{\uparrow})}>\Bigr ]
\end{eqnarray}
the first term is the conventional quark model
result. The $a_8$, $a_{10}$ terms are the corrections
coming from the scalar sea, $b_{1,8,10}$, $c_8$ and
$b_8c_8$ terms are from the vector sea and the $d_8$
term is from the tensor sea.

If operator $\hat O$ has a form like
$\hat O$=${\sum}_i{\hat O}_f^i{\sigma}_z^i$ where
${\hat O}_f^i$ depends only on the flavor of the ith
quark and ${\sigma}_z^i$ is the spin projection
(z direction) operator of ith quark, from (3.1) we
obtain

\begin{eqnarray}
<{\Phi}_{1/2}^{({\uparrow})}| {\hat O}|
 {\Phi}_{1/2}^{({\uparrow})}>&=&{1\over {N^2}}\Bigl [
a\sum_i[<O_f^i>^{\lambda\lambda}<{\sigma}_z^i>^
{\lambda\uparrow\lambda\uparrow}+
<O_f^i>^{\rho\rho}<{\sigma}_z^i>^
{\rho\uparrow\rho\uparrow}\nonumber \\
&&+2<O_f^i>^{\lambda\rho}<{\sigma}_z^i>^
{\lambda\uparrow\rho\uparrow}]\nonumber \\
&&+b\sum_i(<O_f^i>^{\lambda\lambda}+
<O_f^i>^{\rho\rho})(<{\sigma}_z^i>^
{\lambda\uparrow\lambda\uparrow}+
<{\sigma}_z^i>^{\rho\uparrow\rho\uparrow})\nonumber \\
&&+c\sum_i[<O_f^i>^{\lambda\lambda}
<{\sigma}_z^i>^{\rho\uparrow\rho\uparrow}+
<O_f^i>^{\rho\rho}<{\sigma}_z^i>^
{\lambda\uparrow\lambda\uparrow}\nonumber \\
&&-2<O_f^i>^{\lambda\rho}<{\sigma}_z^i>^
{\lambda\uparrow\rho\uparrow}]\nonumber \\
&&+d[\sum_i<O_f^i>^{\lambda\lambda}+
\sum_i<O_f^i>^{\rho\rho}]\nonumber \\
&&+e[\sum_i(<O_f^i>^{\rho\rho}-
<O_f^i>^{\lambda\lambda})<{\sigma}_z^i>^
{\lambda\uparrow{3/2}\uparrow}\nonumber \\
&&+2\sum_i<O_f^i>^{\lambda\rho}<{\sigma}_z^i>^
{\rho\uparrow{3/2}\uparrow}]\Bigr ]
\end{eqnarray}
There are only five combinations of seven parameters appear
in (3.2):

\begin{equation}
$$a={1\over 2}(1-{{b_1^2}\over 3})\ ,\quad
b={1\over 4}(a_8^2-{{b_8^2}\over 3})\ ,
\quad c={1\over 2}(a_{10}^2-{{b_{10}^2}\over 3})
               $$
\end{equation}

\begin{equation}
$$d={1\over {18}}(5c_8^2-3d_8^2)\ ,\quad
e={{\sqrt 2}\over 3}b_8c_8              $$
\end{equation}
and
$<O_f^i>^{\lambda\lambda}\equiv <{\phi}^{\lambda}| O_f^i
| {\phi}^{\lambda}>$,
$<{\sigma}_z^i>^{\lambda\uparrow\lambda\uparrow}
\equiv<{\chi}^{\lambda\uparrow}| {\sigma}_z^i|
{\chi}^{\lambda\uparrow}>$.
Similar notations are used for
$<O_f^i>^{\rho\rho}$, $<{\sigma}_z^i>^{\rho\uparrow\rho\uparrow}$ etc.
All matrix elements for octet baryons are listed in
appendix 1.

For magnetic moments, $O_f^i={e^i}/{2m_i}$ (i=u, d, s).
The baryon magnetic moments can be expressed in terms of
the quark magnetic moments (${\mu}_u$, ${\mu}_d$, ${\mu}_s$)
and two parameters $\alpha$ and $\beta$ as follows

\begin{equation}
{\mu}_p=3({\mu}_u\alpha-{\mu}_d\beta), \quad
{\mu}_n=3({\mu}_d\alpha-{\mu}_u\beta)
\end{equation}

\begin{equation}
{\mu}_{\Lambda}={1\over 2}(\alpha-4\beta)({\mu}_u+{\mu}_d)
+(2\alpha+\beta){\mu}_s
\end{equation}

\begin{equation}
{\mu}_{{\Sigma}^+}=3({\mu}_u\alpha-{\mu}_s\beta), \quad
{\mu}_{{\Sigma}^-}=3({\mu}_d\alpha-{\mu}_s\beta), \quad
{\mu}_{{\Sigma}^0}={1\over 2}({\mu}_{{\Sigma}^+}+{\mu}_{{\Sigma}^-}),
\end{equation}

\begin{equation}
{\mu}_{{\Xi}^0}=3({\mu}_s\alpha-{\mu}_u\beta), \quad
{\mu}_{{\Xi}^-}=3({\mu}_s\alpha-{\mu}_d\beta)
\end{equation}
Also, the transition moment

\begin{equation}
$${\mu}_{\Sigma\Lambda}=-{{\sqrt 3}\over 2}(\alpha+2\beta)
({\mu}_u-{\mu}_d),                   $$
\end{equation}
where ${\mu}_{q}={e}/{2m_q}$ ($q=u, d, s$) and

\begin{equation}
$$\alpha={1\over {N^2}}({4\over 9})(2a+2b+3d+{\sqrt 2}e)
                                   $$
\end{equation}

\begin{equation}
$$\beta={1\over {N^2}}({1\over 9})(2a-4b-6c-6d+4{\sqrt 2}e)
                                   $$
\end{equation}

One may ask why the seven parameters ($a_i$, $b_i$ etc.) in
the wave function contribute only through the combinations
given by $\alpha$ and $\beta$. The physical reason is that
$\alpha$ and $\beta$ are connected with the number of spin-up
(n($q_{\uparrow}$)) and spin-down (n($q_{\downarrow}$)) quarks in
the spin-up proton. If, $\Delta q\equiv n(q_{\uparrow})-
n(q_{\downarrow})+n({\bar q}_{\uparrow})-
n({\bar q}_{\downarrow})$, $q=u, d, s$ then $\Delta u=3\alpha$
and $\Delta d=-3\beta$. This can be directly checked from the
wave function given in (2.15). Also, as there are no explicit
antiquarks or s-quarks in the wave function,
$n({\bar q}_{\uparrow})-n({\bar q}_{\downarrow})$=0 and
$\Delta s$=0. Further, because of in-built flavour SU(3)
symmetry in the wave function $\alpha$ and $\beta$
determine the other magnetic moments. If there is no sea
contribution, $2a=1$ and $b=c=d=e=0$, then $\alpha=4/9$
and $\beta=1/9$, and the simplest quark model result is
reproduced\cite{pdg92}. A class of models\cite{bartelski90,karl92}
have been recently considered in which the magnetic moments have
been expressed in terms of ${\mu}_q$ and $\Delta q$
($q=u, d, s$) without giving an explicit wave function.
Their expressions reduce to ours on putting $\Delta u=3\alpha$,
$\Delta d=-3\beta$ and $\Delta s=0$ (see Ref.\cite{bartelski90,karl92}).

At first sight, (3.5)$-$(3.9) seem to contain five parameters
${\mu}_q$ ($q=u, d, s$), $\alpha$ and $\beta$. However, as
these always appear as products there are only four effective
parameters which we take to be ${\tilde U}\equiv 3\alpha{\mu}_d$,
${\tilde D}\equiv -3\beta{\mu}_d$, $2p\equiv -{\mu}_u/{\mu}_d>0$ and
$r\equiv {\mu}_s/{\mu}_d>0$. The numerical results for this
four-parameter fit to the magnetic moments are discussed later.
Here we note the following four relations or sum rules
between the eight magnetic moments:

\begin{equation}
$$(4.71)\  {\mu}_{p}-{\mu}_{n}={\mu}_{{\Sigma}^+}
-{\mu}_{{\Sigma}^-}-({\mu}_{{\Xi}^0}-{\mu}_{{\Xi}^-})\
(4.15\pm 0.07)$$
\end{equation}

\begin{equation}
$$(3.68\pm 0.02)\ -6{\mu}_{\Lambda}=({\mu}_{{\Sigma}^+}
+{\mu}_{{\Sigma}^-})-2({\mu}_p+{\mu}_n+{\mu}_{{\Xi}^0}
+{\mu}_{{\Xi}^-})\ (3.36\pm 0.09)$$
\end{equation}

\begin{equation}
$$(3.42\pm 0.26)\ ({\mu}^2_{{\Sigma}^+}-{\mu}^2_{{\Sigma}^-})-
({\mu}^2_{{\Xi}^0}-{\mu}^2_{{\Xi}^-})
={\mu}^2_p-{\mu}^2_n\ (4.14) $$
\end{equation}

\begin{equation}
$$(5.58\pm 0.28)\ -2{\sqrt 3}{\mu}_{\Sigma\Lambda}=
2({\mu}_p-{\mu}_n)-({\mu}_{{\Sigma}^+}
-{\mu}_{{\Sigma}^-})\  (5.83\pm 0.06) $$
\end{equation}

The value of the two sides taken from data\cite{pdg92} are
shown in parenthesis. The three sum rules in (3.12)$-$(3.14)
are not new and hold in the class of models with $\Delta s\neq 0$
referred to above. A discussion of why they are poorly
satisfied can be found in Ref.\cite{bartelski90}.
The sum rule in (3.15), a consequence of the $4-$parameter
model, is surprisingly well satisfied.

The simpler case with three effective parameters
${\mu}_0\alpha$, ${\mu}_0\beta$ and $r$ (${\mu}_0\equiv
e/2m_u$) is of interest since it makes the natural
assumption $m_u=m_d$ or ${\mu}_u=-2{\mu}_d$. This implies
an additional sum rule (apart from the (3.12)$-$(3.15))

\begin{equation}
$$(1.61\pm 0.08)\ \ {\mu}_{\Sigma\Lambda}={{\sqrt 3}\over 2}{\mu}_n
\ \ (1.66)                                     $$
\end{equation}
which is quite well satisfied\cite{lichtenberg77,franklin84}.

The important point to note is that because of the sea
contribution $\alpha$ and $\beta$ are free parameters
and not restricted to the simple quark model value.
Finally, we note that the failure of the data to satisfy
the relations (3.12)$-$(3.14) implies that one can only
obtain, at best, an approximate fit to the magnetic
moment data in all the above cases.

\section{WEAK DECAY CONSTANTS AND SPIN DISTRIBUTIONS}

For the weak decay constant $(g_A/g_V)$, $O_f^i=2{I_3^i}$
and we obtain

\begin{eqnarray}
(g_A/g_V)_{n\rightarrow p}&=&{1\over {N^2}}({5\over 3})(
2a+{4\over 5}b-{6\over 5}c+{6\over 5}d+{8\over 5}{\sqrt 2}e)\nonumber \\
&=&3(\alpha+\beta)
\end{eqnarray}

\begin{eqnarray}
(g_A/g_V)_{{\Xi}^-\rightarrow {\Xi}^o}&=&{1\over {N^2}}
(-{1\over 3})(2a-4b-6c-6d+4{\sqrt 2}e)\nonumber \\
&=&-3\beta
\end{eqnarray}
Using (3.5) and (4.1), (4.2) we obtain

\begin{equation}
$$  ({\mu}_{\Xi^0}-{\mu}_{\Xi^-})/
({\mu}_p-{\mu}_n)=
(g_A/g_V)_{{\Xi}^-\rightarrow {\Xi}^o}/
(g_A/g_V)_{n\rightarrow p} $$
\end{equation}
Note that the relation Eq. (4.3) continues to hold in
models\cite{bartelski90,karl92} with $\Delta s\neq 0$ mentioned
above. For the 3-parameter model (i.e. with ${\mu}_u=-2{\mu}_d$)
in addition to (4.3), one obtains

\begin{equation}
$$  ({\mu}_p+2{\mu}_n)/({\mu}_p-{\mu}_n)=
(g_A/g_V)_{{\Xi}^-\rightarrow {\Xi}^o}/
(g_A/g_V)_{n\rightarrow p} $$
\end{equation}
The relations (4.3) and (4.4) cannot be checked directly with
data as ${(g_A/g_V)_{{\Xi}^-\rightarrow {\Xi}^o}}$ is not
measured. However, we can predict (see Table 1) the
${(g_A/g_V)}$ for various semi-leptonic decays since
they can be expressed, using flavour SU(3) symmetry,
in terms of $F$ and $D$ or $\alpha$ and $\beta$.

In fact $(g_A/g_V)_{n\rightarrow p}=F+D$ and
$(g_A/g_V)_{{\Xi}^-\rightarrow {\Xi}^o}=F-D$, from
(4.1) and (4.2) we have

\begin{equation}
$$F=3\alpha/2\ ;\qquad D=3(\alpha+2\beta)/2
\ ;\qquad {F/D}={\alpha}/(\alpha+2\beta)
                                   $$
\end{equation}
It is easy to verify that when there is no sea contribution
({\it i.e.} $a_{8,10}=b_{1,8,10}=c_8=d_8=0$) and ${\mu}_u=
-2{\mu}_d$, the standard SU(6) quark model results, {\it e.g.}
${\mu}_n/{\mu}_p=-2/3$, $(g_A/g_V)_{n\rightarrow p}=5/3$
and $F/D=2/3$ follow.

For spin distributions in the proton and neutron, we have

\begin{equation}
$$I_1^p={1\over 2}<\sum_ie_i^2{\sigma}_z^i>_p
={1\over {3N^2}}({5\over 3}a+2b+{c\over 3}+3d+
{2\over 3}{\sqrt 2}e)
 $$
\end{equation}

\begin{equation}
$$I_1^n={1\over 2}<\sum_ie_i^2{\sigma}_z^i>_n
={1\over {3N^2}}({4\over 3}b+{4\over 3}c+2d-
{2\over 3}{\sqrt 2}e)
 $$
\end{equation}
where $I_1^p\equiv \int g_1^p(x)dx$ etc. Similarly, one can
obtain

\begin{equation}
$$I_1^{\Lambda}={1\over 2}
<\sum_ie_i^2{\sigma}_z^i>_{\Lambda}
={1\over {3N^2}}({1\over 3}a+{4\over 3}b+c+2d-
{2\over 3}{\sqrt 2}e)
 $$
\end{equation}
Using the parameters $\alpha$ and $\beta$, they are

\begin{equation}
$$I_1^p={1\over 6}(4\alpha-\beta)\ ;\qquad
I_1^n={1\over 6}(\alpha-4\beta)\ ;\qquad
I_1^{\Lambda}={1\over 4}(\alpha-2\beta)      $$
\end{equation}
One can see that the standard SU(6) result gives
$\int g_1^p(x)dx=5/18$, $\int g_1^n(x)dx=0$ and
$\int g_1^{\Lambda}(x)dx=1/18$.
Including the sea contributions, however, $\int g_1^p(x)dx$,
$\int g_1^{\Lambda}(x)dx$ could be different from their
SU(6) value and also $\int g_1^n(x)dx$ could be nonzero.
One can verify, however, that the Bjorken sum rule is still
satisfied

\begin{equation}
$$\int [g_1^p(x)-g_1^n(x)]dx={1\over 6}
(g_A/g_V)_{n\rightarrow p}                $$
\end{equation}
In addition, we have

\begin{equation}
$$\int [g_1^p(x)-g_1^{\Lambda}(x)]dx={1\over {12}}
[(g_A/g_V)_{n\rightarrow p}+
(g_A/g_V)_{\Lambda\rightarrow p}]       $$
\end{equation}
In our model, $\int_0^1g_1^{\Lambda}(x)dx$ will be less than
its SU(6) value if sea contributions are taken into account
(see Table 1). It is interesting to note that an experiment
to measure the spin structure function of the $\Lambda$-particle
has been suggested recently\cite{burkhardt93}.

\section{DISCUSSION OF THE SEA CONTRIBUTION}

For simplicity, we consider the case when the magnetic
moments are given by three parameters $\alpha$, $\beta$
and $r$ ({\it i.e.} put ${\mu}_u=-2{\mu}_d$ in (3.5)$-$(3.9)).
The discussion for the case when ${\mu}_u\neq -2{\mu}_d$
can be carried out on similar lines and suggests that
$-{\mu}_u/2{\mu}_d<1$ for both pure scalar and vector sea.

\subsection{Scalar sea}
If sea spin is zero, $a_8\neq 0$ and
$a_{10}\neq 0$, but $b_1=b_8=b_{10}=c_8=d_8=0$, one
obtains

\begin{equation}
$${\mu}_n/{\mu}_p=(-{2\over 3}){{1-a_{10}^2}\over
{1+{1\over 3}(a_8^2-a_{10}^2)}}\simeq
(-{2\over 3})(1-{1\over 3}a_8^2-{2\over 3}a_{10}^2+...)
$$
\end{equation}
since $0\leq a_8^2, a_{10}^2\leq 1$. It is obvious that
the contribution from the scalar sea leads to a wrong
correction to the ratio of neutron and proton magnetic
moments. For $F/D$ ratio, one obtains

\begin{equation}
$$F/D={2\over 3}(1+{1\over 2}a_8^2+a_{10}^2+...)
$$
\end{equation}
which also disagrees with the data. Furthermore, for scalar
sea, the first moment of the neutron spin structure function

\begin{equation}
$$\int g_1^n(x)dx={1\over {9N^2}}(a_8^2+2a_{10}^2)$$
\end{equation}
is positive which seems to contradict the negative
value indicated by the earlier analysis of the EMC
result\cite{ashman89} and the latest data given by the SMC
Collaboration\cite{adeva93}.

\subsection{Vector sea}
We first look at a special vector sea as discussed
in Ref.\cite{li91}. Assuming $a_{8,10}$=$b_{1,10}$=$d_8$=0
and $b_{8}$=$c_{8}$=$-\delta$, it is easy to see,
from (3.5) that

\begin{equation}
$${\mu}_p={\mu}_0{{1+{4\over 3}{\delta}^2}\over
{1+2{\delta}^2}}\ , \quad
{\mu}_n={\mu}_0(-{2\over 3}){{1+{4\over 3}
{\delta}^2}\over {1+2{\delta}^2}}
                          $$
\end{equation}
where ${\mu}_0=e/2m_u$. Hence the relation
${\mu}_n/{\mu}_p=-2/3$ is preserved
as given in \cite{li91}. Similarly, from (4.1) and (4.2), we
have

\begin{equation}
$$(g_A/g_V)_{n\rightarrow p}=({5\over 3})
{{1+{4\over 3}{\delta}^2}\over {1+2{\delta}^2}}
                              $$
\end{equation}

\begin{equation}
$$(g_A/g_V)_{{\Xi}^-\rightarrow {\Xi}^o}=
(-{1\over 3}){{1+{4\over 3}{\delta}^2}\over
{1+2{\delta}^2}}
                              $$
\end{equation}
one can see that the conventional SU(6) result
$(g_A/g_V)_{n\rightarrow p}/
(g_A/g_V)_{{\Xi}^-\rightarrow {\Xi}^o}=-5$
is also preserved. However, using parameter $\delta
=-0.35$ given in \cite{li91}, we obtain
$(g_A/g_V)_{n\rightarrow p}$=1.727,
which is inconsistent with the data\cite{pdg92}
$(g_A/g_V)_{n\rightarrow p}$=1.257$\pm$0.003.
This disagreement is not unexpected. Because the
perturbative calculation of the mixing parameters
and its result $b_8=c_8=-\delta$ are questionable.
It is obvious that the nonperturbative effects,
which are dominant in the low energy region, would
change the relative weight of these mixing parameters
significantly. Therefore, we prefer to discuss a more
general vector sea and to look if there is another
appropriate parameter set, in which the nonperturbative
and perturbative effects are taking into account, can
lead to a better agreement with the low energy
baryon properties. We will show below that this
parameter set not only gives a right
modification to the ratio ${\mu}_n/{\mu}_p$
but also gives a very good result for axial coupling
constants.

As we mentioned above, the mixing parameters basically
come from the nonperturbative interactions between
quarks and gluons. Hence we do not attempt to calculate
these parameters, but rather estimate them by the
required agreement with the low energy data. Before
doing this, we give some arguments as motivations
for choosing the parameters. Since the sea basically
comes from the emission of virtual gluons, the $b_8$
term would be dominant and we would expect

\begin{equation}
$$b_1^2, b_{10}^2\ ({\rm two}-{\rm gluon}\
{\rm sea})<<b_8^2\ ({\rm one}-{\rm gluon}\ {\rm sea})
                                       $$
\end{equation}
The $c_8$ term is expected to be small due to another
reason

\begin{equation}
$$c_8^2\ ({\rm quark}\ {\rm spin}-{\rm flip})<<
b_8^2\ ({\rm quark}\ {\rm spin}-{\rm nonflip})
                                       $$
\end{equation}
The scalar sea $a_8$ and $a_{10}$ terms are
expected to be also small because they can only
come from the two-gluon sea. The tensor sea ($d_8$)
term comes from two-gluon sea and quark spin-flip
process, hence it should be highly suppressed.
Assuming no scalar and tensor sea contribution
and neglecting the $c_8^2$ term (since $c_8^2<<
b_8^2$), we have

\begin{equation}
$${\mu}_n/{\mu}_p=(-2/3){{1-{1\over 3}b_{1}^2}\over
{1-{1\over 3}b_1^2-{1\over 9}b_{8}^2}}\simeq
(-{2\over 3})(1+{1\over 9}b_8^2)
$$
\end{equation}
thus the sea contribution gives a correction in the
right direction.

\section{NUMERICAL RESULTS}

To obtain numerical results, we use the data on magnetic
moments and weak decay coupling constants to determine
the parameters. In particular, the values of $\alpha$
and $\beta$ so obtained should be reproducible by choice
of the seven basic parameters $a_8$, $a_{10}$, $b_1$,
$b_8$ etc. which determine the sea contribution. It is
clear from (3.10) and (3.11) that there are many ways of
choosing $a_8$ etc. to give the same $\alpha$ and $\beta$.
However, guided by the qualitative discussion of section
V, we will assume the sea is mainly vector with a small
scalar component. The tensor sea is neglected ($d_8$=0).
We shall see that the parameters ($b_8$, $c_8$ etc.),
which determine the contribution of such a sea to the
baryon structure, can be chosen to give the $\alpha$
and $\beta$ determined from the data.

\subsection{Four-parameter fit}
The magnetic moments in (3.5)$-$(3.9) are given in terms
of four effective parameters ${\tilde U}\equiv 3\alpha{\mu}_d$,
${\tilde D}\equiv -3\beta{\mu}_d$, $2p\equiv -{\mu}_u/{\mu}_d>0$
and $r\equiv {\mu}_s/{\mu}_d>0$. Using ${\mu}_p$,
${\mu}_n$, ${\mu}_{\Sigma, \Lambda}$ as inputs one can
directly determine ${\tilde U}=-1.348$, ${\tilde D}=0.306$
and $p=0.922$ as these do not involve the parameter $r$.
The value of ${\mu}_{\Lambda}$ is used as input to fix
$r=0.6255$. Knowledge of the ratio ${\alpha}/{\beta}$=4.406
immediately predicts (see (4.5))

\begin{equation}
$$ F/D=0.6878
$$
\end{equation}
A more realistic model with a small $\Delta s\neq 0$
could easily modify this value. Note that in the models
of \cite{karl92} and \cite{bartelski90} with extra parameter
($\Delta s$) they obtain 0.726 and
0.585 for this ratio. To separate out the parameters
${\alpha}$ and ${\beta}$, we use the axial coupling
constant data to obtain

\begin{equation}
$${\alpha}=0.3415,\quad {\beta}=0.0775
$$
\end{equation}
The values obtained for the quark magnetic moments (in
nuclear magnetons ${\mu}_N$) are

\begin{equation}
$${\mu}_u=2.428,\ \ {\mu}_d=-1.316,\ \ {\mu}_s=-0.823 $$
\end{equation}
A choice of sea parameters which reproduce the parameters
${\alpha}$ and ${\beta}$ given in (6.2) are $b_1^2=0.0039$,
$b_8^2=0.22$, $c_8^2=0.027$ (for vector sea) and
$a_{10}^2=0.0975$ (for scalar sea) with $b_8c_8>0$.

The values obtained for magnetic moments and other
quantities are displayed in column 4 of Table I. It
can be seen that the fit to the magnetic moments and
the axial coupling constants is quite reasonable except
for ${\mu}_{{\Sigma}^+}$. For the quark spin distributions
our calculation suggests a small non-zero negative value
for $\int_0^1g_1^n(x)dx$, however, the result for
$\int_0^1g_1^p(x)dx$=0.2147 is much larger than the
experimental value\cite{ashman89} of $0.126\pm 0.018$. One must
note, however, that the EMC experiment gives this value
for $<Q^2>$=10.7 (GeV/c)$^2$ and this can be very different
from the very low $Q^2$ result predicted by our $q^3$+sea
model.

\subsection{Three-parameter fit}
The natural assumption $m_u=m_d$ implies the relation
${\mu}_u=-2{\mu}_d$. Implementing relation in (3.5)$-$(3.9)
gives ${\mu}_p={\mu}_0(2\alpha+\beta)$,
${\mu}_n=-{\mu}_0(\alpha+2\beta)$ etc. where ${\mu}_0
\equiv e/2m_u$. The magnetic moments are given in terms of
three effective parameters ${\mu}_0\alpha$, ${\mu}_0\beta$,
and $r$. Guided by 4-parameter fit we choose sea parameters
similar to that case, namely  $b_1^2=0.1$, $b_8^2=0.22$,
$c_8^2=0.027$ and $a_{10}^2=0.02$ with $b_8c_8>0$. Basically
we have enhanced the vector sea with a larger value of $b_1$
and reduced the scalar sea with a smaller value of $a_{10}$.
This choice immediately gives $\alpha=0.3264$ and $\beta=
0.0927$. Using ${\mu}_p$ and ${\mu}_{\Lambda}$ as inputs
then determines ${\mu}_0=3.7465{\mu}_N$ and $r=0.6286$.
The results of magnetic moments etc. are listed in column
5 of Table I. Since the ratio $\alpha/\beta=3.521$ one obtains
\begin{equation}
$$    F/D=0.6380                     $$
\end{equation}
which is fairly close to the experimental value\cite{bourquin83}.
Since
$F/D$ increases monotonically with increasing $\alpha/\beta$,
for the simple quark model ($\alpha/\beta=4$) the value of
$F/D$=2/3 lies between those in (6.1) and (6.4). The results for
quark spin distributions are similar to the $4-$parameter case.

For comparison, in column 3 of Table I the results for
the simple quark model are given. In this case there is no
sea contribution and the baryons are given by standard $q^3$
wave function which fixes $\alpha=4/9$ and $\beta=1/9$. The
magnetic moments are given in terms of $3-$parameters
${\mu}_u$, ${\mu}_d$ and ${\mu}_s$. This fit with ${\mu}_p$,
${\mu}_n$ and ${\mu}_{\Lambda}$ as inputs gives
${\mu}_u/(-2{\mu}_d)=p$=0.953 and $r={\mu}_s/{\mu}_u$=0.63.
We have used the same inputs in all three cases for a
meaningful comparison. From Table I one can see that the
$4-$parameter gives a somewhat better overall fit.

\section{SUMMARY}

In summary, we have suggested a general formalism to treat
a baryon as a composite system of $q^3$ plus a flavorless
sea. The modifications of the different properties of spin
1/2 baryon, by the sea, are given. Numerical fits to the
individual magnetic moments, $\Sigma\Lambda-$transition
moment and axial weak coupling constants for the baryon
octet have been obtained. These results seem to favour
a dominantly vector sea.

It should be noted that our results and conclusions are
subject to the following points:
(i) the sea and the $3-$quarks are considered to be in a
relative $S-$state, possible higher angular momentum
states have been neglected; (ii) the sea is assumed
to be flavorless and has been specified only by its
total quantum numbers; (iii) further, modification of baryon
wave function is needed to have non$-$zero $\Delta s$ in
the nucleon; (iv) relativistic corrections have been
neglected although the internal motion of the light
quarks in the baryon is expected to be relativistic;
(v) all calculations have been performed in the baryon
rest frame. This may be reasonable for the magnetic
moments and the weak decay constants, but may not
be appropriate for comparing the spin distribution
calculated by us (at low $Q^2-$scale) with the EMC
data at much high momentum transfer. All these points
need to be considered in future work to fully understand
baryon structure.

\acknowledgments
The authors thank J. S. McCarthy and P. K. Kabir for their
useful comments and suggestions. V. G. would like to thank
J. S. McCarthy for his warm hospitality at the Institute
of Nuclear and Particle Physics when this work was started.
X. Song was supported by the US Department of Energy and
the Commonwealth Center for Nuclear and High Energy Physics,
Virginia, USA.

\vfill\eject

\appendix{Appendix. Matrix Elements for Different Operators}

(i) Spin Projection Operator

\begin{equation}
$$<{\sigma}_z^{(1)}>^{\lambda\uparrow\lambda\uparrow}=
<{\sigma}_z^{(2)}>^{\lambda\uparrow\lambda\uparrow}=2/3
\ ;\qquad
<{\sigma}_z^{(3)}>^{\lambda\uparrow\lambda\uparrow}=-1/3
$$
\end{equation}

\begin{equation}
$$<{\sigma}_z^{(1)}>^{\rho\uparrow\rho\uparrow}=
<{\sigma}_z^{(2)}>^{\rho\uparrow\rho\uparrow}=0
\ ;\qquad
<{\sigma}_z^{(3)}>^{\rho\uparrow\rho\uparrow}=1
$$
\end{equation}

\begin{equation}
$$<{\sigma}_z^{(1)}>^{\lambda\uparrow\rho\uparrow}=
-<{\sigma}_z^{(2)}>^{\lambda\uparrow\rho\uparrow}=1/{\sqrt 3}
\ ;\qquad
<{\sigma}_z^{(3)}>^{\lambda\uparrow\rho\uparrow}=0
$$
\end{equation}
It is easy to see that the matrix elements in (1) and
(2) satisfy

\begin{equation}
$$ \sum_{i=1}^3<{\sigma}_z^{(i)}>^{\lambda\uparrow\lambda\uparrow}=
\sum_{i=1}^3<{\sigma}_z^{(i)}>^{\rho\uparrow\rho\uparrow}=1
$$
\end{equation}
In addition,

\begin{equation}
$$<{\sigma}_z^{(1)}>^{\lambda\uparrow 3/2\uparrow}=
<{\sigma}_z^{(2)}>^{\lambda\uparrow 3/2\uparrow}=-{\sqrt 2}/3
\ ;\qquad
<{\sigma}_z^{(3)}>^{\lambda\uparrow 3/2\uparrow}=2{\sqrt 2}/3
$$
\end{equation}

\begin{equation}
$$<{\sigma}_z^{(1)}>^{\rho\uparrow 3/2\uparrow}=
-<{\sigma}_z^{(2)}>^{\rho\uparrow 3/2\uparrow}={\sqrt {2/3}}
\ ;\qquad
<{\sigma}_z^{(3)}>^{\rho\uparrow 3/2\uparrow}=0
$$
\end{equation}

The matrix elements in (3), (5) and (6) satisfy

\begin{equation}
$$ \sum_{i=1}^3<{\sigma}_z^{(i)}>^{\lambda\uparrow\rho\uparrow}=
 \sum_{i=1}^3<{\sigma}_z^{(i)}>^{\lambda\uparrow 3/2\uparrow}=
 \sum_{i=1}^3<{\sigma}_z^{(i)}>^{\rho\uparrow 3/2\uparrow}=0
                                $$
\end{equation}

(ii) Isospin Projection Operator

For the proton, we have

\begin{equation}
$$<{I}_3^{(1)}>_p^{\lambda\lambda}=
<{I}_3^{(2)}>_p^{\lambda\lambda}=1/3
\ ;\qquad
<{I}_3^{(3)}>_p^{\lambda\lambda}=-1/6
$$
\end{equation}

\begin{equation}
$$<{I}_3^{(1)}>_p^{\rho\rho}=<{I}_3^{(2)}>^{\rho\rho}=0
\ ;\qquad
<{I}_3^{(3)}>_p^{\rho\rho}=1/2
$$
\end{equation}

\begin{equation}
$$<{I}_3^{(1)}>_p^{\lambda\rho}=
-<{I}_3^{(2)}>_p^{\lambda\rho}=1/{2\sqrt 3}
\ ;\qquad
<{I}_3^{(3)}>_p^{\lambda\rho}=0
$$
\end{equation}
for the neutron, all matrix elements get an opposite sign.
For $\Sigma^+$-hyperon, we have

\begin{equation}
$$<{I}_3^{(1)}>_{\Sigma^+}^{\lambda\lambda}=
<{I}_3^{(2)}>_{\Sigma^+}^{\lambda\lambda}=5/12
\ ;\qquad
<{I}_3^{(3)}>_{\Sigma^+}^{\lambda\lambda}=1/6
$$
\end{equation}

\begin{equation}
$$<{I}_3^{(1)}>_{\Sigma^+}^{\rho\rho}
=<{I}_3^{(2)}>_{\Sigma^+}^{\rho\rho}=1/4
\ ;\qquad
<{I}_3^{(3)}>_{\Sigma^+}^{\rho\rho}=1/2
$$
\end{equation}

\begin{equation}
$$<{I}_3^{(1)}>_{\Sigma^+}^{\lambda\rho}=
-<{I}_3^{(2)}>_{\Sigma^+}^{\lambda\rho}=1/{4\sqrt 3}
\ ;\qquad
<{I}_3^{(3)}>_{\Sigma^+}^{\lambda\rho}=0
$$
\end{equation}
similarly, for ${\Sigma^-}$ the matrix elements reverse their signs.
For $\Xi^0$-hyperon, we have

\begin{equation}
$$<{I}_3^{(1)}>_{\Xi^0}^{\lambda\lambda}=
<{I}_3^{(2)}>_{\Xi^0}^{\lambda\lambda}=1/12
\ ;\qquad
<{I}_3^{(3)}>_{\Xi^0}^{\lambda\lambda}=1/3
$$
\end{equation}

\begin{equation}
$$<{I}_3^{(1)}>_{\Xi^0}^{\rho\rho}=<{I}_3^{(2)}>_{\Xi^0}^{\rho\rho}=1/4
\ ;\qquad
<{I}_3^{(3)}>_{\Xi^0}^{\rho\rho}=0
$$
\end{equation}

\begin{equation}
$$<{I}_3^{(1)}>_{\Xi^0}^{\lambda\rho}=
-<{I}_3^{(2)}>_{\Xi^0}^{\lambda\rho}=-1/{4\sqrt 3}
\ ;\qquad
<{I}_3^{(3)}>_{\Xi^0}^{\lambda\rho}=0
$$
\end{equation}
for ${\Xi^-}$, all matrix elements reverse their signs. Finally,
all isospin matrix elements for $\Lambda$ and $\Sigma^0$ hyperons are zero.

(iii) Charge Operator With Symmetry Breaking Effect

For the proton, we have

\begin{equation}
$$<{e}^{(1)}{m\over {m_1}}>_p^{\lambda\lambda}=
<{e}^{(2)}{m\over {m_2}}>_p^{\lambda\lambda}=1/2
\ ;\qquad
<{e}^{(3)}{m\over {m_3}}>_p^{\lambda\lambda}=0
$$
\end{equation}

\begin{equation}
$$<{e}^{(1)}{m\over {m_1}}>_p^{\rho\rho}=
<{e}^{(2)}{m\over {m_2}}>_p^{\rho\rho}=1/6
\ ;\qquad
<{e}^{(3)}{m\over {m_3}}>_p^{\rho\rho}=2/3
$$
\end{equation}

\begin{equation}
$$<{e}^{(1)}{m\over {m_1}}>_p^{\lambda\rho}=
-<{e}^{(2)}{m\over {m_2}}>_p^{\lambda\rho}=1/{2\sqrt 3}
\ ;\qquad
<{e}^{(3)}{m\over {m_3}}>_p^{\lambda\rho}=0
$$
\end{equation}
where $m=m_u=m_d$. We note that the matrix element
$<{e}^{(3)}{m\over {m_3}}>^{\lambda\rho}$ vanishes for all
octet baryons.

For the neutron, the matrix elements in
(19) reverse their signs. But in (18) the first two
matrix elements $\bf do$ $\bf not$ change the sign, i.e.
$<{e}^{(i)}{m\over {m_i}}>_n^{\rho\rho}=
<{e}^{(i)}{m\over {m_i}}>_p^{\rho\rho}$ (i=1,2) and the
third one becomes $<{e}^{(3)}{m\over {m_3}}>_n^{\rho\rho}=-1/3$.
For the neutron matrix elements in (17), we have

\begin{equation}
$$<{e}^{(1)}{m\over {m_1}}>_n^{\lambda\lambda}=
<{e}^{(2)}{m\over {m_2}}>_n^{\lambda\lambda}=-1/6
\ ;\qquad
<{e}^{(3)}{m\over {m_3}}>_n^{\lambda\lambda}=1/3
$$
\end{equation}
For ${\Sigma^+}$, we obtain

\begin{equation}
$$<{e}^{(1)}{m\over {m_1}}>_{\Sigma^+}^{\lambda\lambda}=
<{e}^{(2)}{m\over {m_2}}>_{\Sigma^+}^{\lambda\lambda}=(10-r)/18
\ ;\qquad
<{e}^{(3)}{m\over {m_3}}>_{\Sigma^+}^{\lambda\lambda}=2(1-r)/9
$$
\end{equation}

\begin{equation}
$$<{e}^{(1)}{m\over {m_1}}>_{\Sigma^+}^{\rho\rho}=
<{e}^{(2)}{m\over {m_2}}>_{\Sigma^+}^{\rho\rho}=(2-r)/6
\ ;\qquad
<{e}^{(3)}{m\over {m_3}}>_{\Sigma^+}^{\rho\rho}=2/3
$$
\end{equation}

\begin{equation}
$$<{e}^{(1)}{m\over {m_1}}>_{\Sigma^+}^{\lambda\rho}=
-<{e}^{(2)}{m\over {m_2}}>_{\Sigma^+}^{\lambda\rho}=(2+r)/{6\sqrt 3}
$$
\end{equation}
while for ${\Sigma^-}$, we have

\begin{equation}
$$<{e}^{(1)}{m\over {m_1}}>_{\Sigma^-}^{\lambda\lambda}=
<{e}^{(2)}{m\over {m_2}}>_{\Sigma^-}^{\lambda\lambda}=-(5+r)/18
\ ;\qquad
<{e}^{(3)}{m\over {m_3}}>_{\Sigma^-}^{\lambda\lambda}=-(1+2r)/9
$$
\end{equation}

\begin{equation}
$$<{e}^{(1)}{m\over {m_1}}>_{\Sigma^-}^{\rho\rho}=
<{e}^{(2)}{m\over {m_2}}>_{\Sigma^-}^{\rho\rho}=-(1+r)/6
\ ;\qquad
<{e}^{(3)}{m\over {m_3}}>_{\Sigma^-}^{\rho\rho}=-1/3
$$
\end{equation}

\begin{equation}
$$<{e}^{(1)}{m\over {m_1}}>_{\Sigma^-}^{\lambda\rho}=
-<{e}^{(2)}{m\over {m_2}}>_{\Sigma^-}^{\lambda\rho}=-(1-r)/{6\sqrt 3}
$$
\end{equation}
For $\Xi^0$, we have

\begin{equation}
$$<{e}^{(1)}{m\over {m_1}}>_{\Xi^0}^{\lambda\lambda}=
<{e}^{(2)}{m\over {m_2}}>_{\Xi^0}^{\lambda\lambda}=(2-5r)/18
\ ;\qquad
<{e}^{(3)}{m\over {m_3}}>_{\Xi^0}^{\lambda\lambda}=(4-r)/9
$$
\end{equation}

\begin{equation}
$$<{e}^{(1)}{m\over {m_1}}>_{\Xi^0}^{\rho\rho}=
<{e}^{(2)}{m\over {m_2}}>_{\Xi^0}^{\rho\rho}=(2-r)/6
\ ;\qquad
<{e}^{(3)}{m\over {m_3}}>_{\Xi^0}^{\rho\rho}=-r/3
$$
\end{equation}

\begin{equation}
$$<{e}^{(1)}{m\over {m_1}}>_{\Xi^0}^{\lambda\rho}=
-<{e}^{(2)}{m\over {m_2}}>_{\Xi^0}^{\lambda\rho}=-(2+r)/{6\sqrt 3}
$$
\end{equation}
and for $\Xi^-$

\begin{equation}
$$<{e}^{(1)}{m\over {m_1}}>_{\Xi^-}^{\lambda\lambda}=
<{e}^{(2)}{m\over {m_2}}>_{\Xi^-}^{\lambda\lambda}=-(1+5r)/18
\ ;\qquad
<{e}^{(3)}{m\over {m_3}}>_{\Xi^-}^{\lambda\lambda}=-(2+r)/9
$$
\end{equation}

\begin{equation}
$$<{e}^{(1)}{m\over {m_1}}>_{\Xi^-}^{\rho\rho}=
<{e}^{(2)}{m\over {m_2}}>_{\Xi^-}^{\rho\rho}=-(1+r)/6
\ ;\qquad
<{e}^{(3)}{m\over {m_3}}>_{\Xi^-}^{\rho\rho}=-r/3
$$
\end{equation}

\begin{equation}
$$<{e}^{(1)}{m\over {m_1}}>_{\Xi^-}^{\lambda\rho}=
-<{e}^{(2)}{m\over {m_2}}>_{\Xi^-}^{\lambda\rho}=(1-r)/{6\sqrt 3}
$$
\end{equation}
For $\Lambda^0$, we obtain

\begin{equation}
$$<{e}^{(1)}{m\over {m_1}}>_{\Lambda^0}^{\lambda\lambda}=
<{e}^{(2)}{m\over {m_2}}>_{\Lambda^0}^{\lambda\lambda}=(1-2r)/12
\ ;\qquad
<{e}^{(3)}{m\over {m_3}}>_{\Lambda^0}^{\lambda\lambda}=1/6
$$
\end{equation}

\begin{equation}
$$<{e}^{(1)}{m\over {m_1}}>_{\Lambda^0}^{\rho\rho}=
<{e}^{(2)}{m\over {m_2}}>_{\Lambda^0}^{\rho\rho}=(5-2r)/36
\ ;\qquad
<{e}^{(3)}{m\over {m_3}}>_{\Lambda^0}^{\rho\rho}=(1-4r)/18
$$
\end{equation}

\begin{equation}
$$<{e}^{(1)}{m\over {m_1}}>_{\Lambda^0}^{\lambda\rho}=
-<{e}^{(2)}{m\over {m_2}}>_{\Lambda^0}^{\lambda\rho}=-(1+2r)/{12\sqrt 3}
$$
\end{equation}
and for $\Sigma^0$, one obtains

\begin{equation}
$$<{e}^{(1)}{m\over {m_1}}>_{\Sigma^0}^{\lambda\lambda}=
<{e}^{(2)}{m\over {m_2}}>_{\Sigma^0}^{\lambda\lambda}=(5-2r)/36
\ ;\qquad
<{e}^{(3)}{m\over {m_3}}>_{\Sigma^0}^{\lambda\lambda}=(1-4r)/18
$$
\end{equation}

\begin{equation}
$$<{e}^{(1)}{m\over {m_1}}>_{\Sigma^0}^{\rho\rho}=
<{e}^{(2)}{m\over {m_2}}>_{\Sigma^0}^{\rho\rho}=(1-2r)/12
\ ;\qquad
<{e}^{(3)}{m\over {m_3}}>_{\Sigma^0}^{\rho\rho}=1/6
$$
\end{equation}

\begin{equation}
$$<{e}^{(1)}{m\over {m_1}}>_{\Sigma^0}^{\lambda\rho}=
-<{e}^{(2)}{m\over {m_2}}>_{\Sigma^0}^{\lambda\rho}=(1+2r)/{12\sqrt 3}
$$
\end{equation}
Finally, for $\Sigma^0\rightarrow\Lambda^0$ transition elements we have

\begin{equation}
$$<{e}^{(1)}{m\over {m_1}}>_{\Sigma^0\rightarrow\Lambda^0}
^{\lambda\lambda}=<{e}^{(2)}{m\over {m_2}}>_{\Sigma^0
\rightarrow\Lambda^0}^{\lambda\lambda}=-1/{4\sqrt 3}
\ ;\qquad
<{e}^{(3)}{m\over {m_3}}>_{\Sigma^0\rightarrow\Lambda^0}
^{\lambda\lambda}=1/{2\sqrt 3}
$$
\end{equation}

\begin{equation}
$$<{e}^{(1)}{m\over {m_1}}>_{\Sigma^0\rightarrow\Lambda^0}
^{\rho\rho}=
<{e}^{(2)}{m\over {m_2}}>_{\Sigma^0\rightarrow\Lambda^0}
^{\rho\rho}=1/{4\sqrt 3}
\ ;\qquad
<{e}^{(3)}{m\over {m_3}}>_{\Sigma^0\rightarrow\Lambda^0}
^{\rho\rho}=-1/{2\sqrt 3}
$$
\end{equation}

\begin{equation}
$$<{e}^{(1)}{m\over {m_1}}>_{\Sigma^0\rightarrow\Lambda^0}
^{\lambda\rho}=
-<{e}^{(2)}{m\over {m_2}}>_{\Sigma^0\rightarrow\Lambda^0}
^{\lambda\rho}=-1/4
$$
\end{equation}

(iv) Charge Square Operator

We only discuss the nucleon case, for the proton we obtain

\begin{equation}
$$<{e}^{(1)^2}>_p^{\lambda\lambda}=
<{e}^{(2)^2}>_p^{\lambda\lambda}=7/18
\ ;\qquad
<{e}^{(3)^2}>_p^{\lambda\lambda}=2/9
$$
\end{equation}

\begin{equation}
$$<{e}^{(1)^2}>_p^{\rho\rho}=<{e}^{(2)^2}>_p^{\rho\rho}=5/18
\ ;\qquad
<{e}^{(3)^2}>_p^{\rho\rho}=4/9
$$
\end{equation}

\begin{equation}
$$<{e}^{(1)^2}>_p^{\lambda\rho}=
-<{e}^{(2)^2}>_p^{\lambda\rho}=1/{6\sqrt 3}
\ ;\qquad
<{e}^{(3)^2}>_p^{\lambda\rho}=0
$$
\end{equation}
while for the neutron, one obtains

\begin{equation}
$$<{e}^{(1)^2}>_n^{\lambda\lambda}=
<{e}^{(2)^2}>_n^{\lambda\lambda}=1/6
\ ;\qquad
<{e}^{(3)^2}>_n^{\lambda\lambda}=1/3
$$
\end{equation}

\begin{equation}
$$<{e}^{(1)^2}>_n^{\rho\rho}=<{e}^{(2)^2}>_n^{\rho\rho}=5/18
\ ;\qquad
<{e}^{(3)^2}>_n^{\rho\rho}=1/9
$$
\end{equation}

\begin{equation}
$$<{e}^{(1)^2}>_n^{\lambda\rho}=
-<{e}^{(2)^2}>_n^{\lambda\rho}=-1/{6\sqrt 3}
\ ;\qquad
<{e}^{(3)^2}>_n^{\lambda\rho}=0
$$
\end{equation}

\begin{table}
\caption{Comparison of the calculated magnetic moments
and axial coupling constants of baryons with data and other models.}
\begin{tabular}{rclcrclcrclcrcl}
 &Baryon      &~~~~&
 &Data$^{16}$ &~~~~&
 &SQM$^a$     &~~~~&
 &Set I$^b$   &~~~~&
 &Set II$^c$  \\
\tableline
&P &~~~~& &2.7928&~~~~&&2.793$^*$&~~~~&&2.7928$^*$&~~~~&&2.793$^*$\\
&n&~~~~& &$-$1.9130&~~~~&&$-$1.913$^*$&~~~~&&$-$1.913$^*$&~~~~&&$-$1.917$^*$\\
&$\Lambda$&~~~~&&$-$0.613$\pm$0.004&~~~~&&$-$0.613$^*$&~~~~&&$-$0.613$^*$
&~~~~&&$-$0.613$^*$\\
&${\Sigma}^0$$\Lambda$&~~~~&&$-$1.61$\pm$0.08&~~~~&&$-$1.63&~~~~&&$-$1.61$^*$
&~~~~&&$-$1.66 \\
& ${\Sigma}^+$&~~~~&& 2.42$\pm$0.05&~~~~&&2.674 &~~~~&& 2.678&~~~~&&2.664\\
& ${\Sigma}^0$&~~~~&&  $-$ &~~~~&&0.791&~~~~&& 0.761&~~~~&&0.830\\
& ${\Sigma}^-$&~~~~&&$-$1.160$\pm$0.025&~~~~&&$-$1.092&~~~~&&$-$1.156
&~~~~&&$-$1.004\\
& ${\Xi}^0$&~~~~&&$-$1.250$\pm$0.014&~~~~&&$-$1.435&~~~~&&$-$1.408
&~~~~&&$-$1.463\\
& ${\Xi}^-$ &~~~~&&$-$0.6507$\pm 0.0025$&~~~~&&$-$0.493&~~~~&&$-$0.537
&~~~~&&$-$0.421\\
& $(g_A/g_V)_{n\rightarrow p}$&~~~~&&1.2573$\pm$ 0.0028&~~~~&& 1.666
&~~~~&&1.2571 &~~~~&&1.2573    \\
& $(g_A/g_V)_{\Lambda\rightarrow p}$&~~~~&&0.718$\pm$ 0.015&~~~~&&1.000
&~~~~&&0.7605&~~~~&&0.7455     \\
& $(g_A/g_V)_{{\Sigma}^-\rightarrow n}$&~~~~&&$-$0.340$\pm$ 0.017&~~~~&
&$-$0.333 &~~~~&&$-$0.2325&~~~~&&$-$0.2781 \\
& $(g_A/g_V)_{{\Xi}^-\rightarrow \Lambda}$&~~~~&&0.25$\pm$ 0.05&~~~~&&
0.333 &~~~~&&0.2640&~~~~&&0.2337 \\
& $(g_A/g_V)_{{\Xi}^-\rightarrow {\Xi}^0}$&~~~~&&$-$ &~~~~&&
$-$0.333 &~~~~&&$-$0.2325&~~~~&&$-$0.2781 \\
& $\int g_1^p(x)dx$&~~~~&&0.126$\pm$0.010$\pm$0.015&~~~~&& 0.278 &~~~~&&
0.2147&~~~~&&0.202\\
& $\int g_1^n(x)dx$&~~~~&& $-$0.08$\pm$0.06$^d$ &~~~~&&0.0 &~~~~&& $-$0.0052
&~~~~&&$-$0.007\\
& $\int g_1^{\Lambda}(x)dx$&~~~~&&$-$&~~~~&&0.0556&~~~~&&0.0466&~~~~&&0.0353\\

\end{tabular}

a) Standard quark model result, e.g. see Ref. 16, VIII.59

b) Four$-$parameter fit

c) Three$-$parameter fit

d) Ref.\cite{adeva93}

*) Inputs

\end{table}

\end{document}